\documentclass[a4paper,twoside]{article}
\newcommand{\affil}[1]{$^{\rm #1}$}
\usepackage[authoryear]{natbib}
\bibpunct{(}{)}{;}{a}{}{,}
\usepackage{graphicx}
\usepackage{url}
\date{} 
\newcommand{\splot}{{\sc S2PLOT}}
\newcommand{\pgplot}{{\sc PGPLOT}}
\newcommand{\stereotwo}{{\sc STEREO2}}
\title{\large\bf\flushleft An Advanced, Three-dimensional Plotting Library for Astronomy}
\author{\parbox{\textwidth}{\flushleft
\vspace{-0.5cm}
{\it David G.\ Barnes\affil{A}, Christopher J.\ Fluke\affil{A,B}, Paul D.\
Bourke\affil{A} and Owen T. Parry\affil{A}}\\
\vspace{0.4cm}
{\small \affil{A}\,Centre for Astrophysics and Supercomputing, Swinburne University of Technology, Hawthorn VIC 3122, Australia}\\
{\small \affil{B}\,Correspondence author. E-mail: cfluke@swin.edu.au}}}
\begin{document}
\twocolumn[
\begin{changemargin}{.8cm}{.5cm}
\begin{minipage}{.9\textwidth}
\vspace{-1cm}
\maketitle
%
%
\small{\bf Abstract:}
We present a new, three-dimensional (3D) plotting library with advanced
features, and support for standard and enhanced display devices.  The
library -- \splot\ -- is written in C and can be used by C, C++ and
FORTRAN programs on GNU/Linux and Apple/OSX systems.  \splot\ draws objects
in a 3D ($x$,$y$,$z$) Cartesian space and the user interactively controls
how this space is rendered at run time.  With a
\pgplot -inspired interface, \splot\ provides astronomers with 
elegant techniques for displaying and exploring 3D data sets directly
from their program code, and the potential to use stereoscopic and
dome display devices.  The \splot\ architecture supports dynamic
geometry and can be used to plot time-evolving data sets, such as
might be produced by simulation codes.  In this paper, we introduce
\splot\ to the astronomical community, describe its potential
applications, and present some example uses of the library.

\medskip{\bf Keywords:} methods: data analysis --- techniques:
miscellaneous --- surveys --- catalogs

\medskip
\medskip
\end{minipage}
\end{changemargin}
]
\small

\section{Introduction}

\subsection{The status quo}

Visualization is a key tool in astronomy for discovery and analysis tool
that  
is used throughout the discipline.  In observational astronomy, data
display applications are used from the observation planning stage
through data collection and reduction phases, to production of figures
for journal articles.  In theoretical astrophysics, data display is an
essential process in comprehending nearly every simulation data set,
and in summarising results for publication.

Most existing astronomy display tools operate in the two-dimensional
(2d) paradigm.  That is, multi-dimensional data is explicitly reduced
to a data set having at most two ``look-up'' coordinates (or indices)
into the data prior to being presented to a display device.   Examples
of 2d display packages widely adopted by the astronomy
community include the {\sc DS9} image display tool\footnote{{\sc DS9}
-- \url{http://hea-www.harvard.edu/RD/ds9}}, the \pgplot\ programming
library\footnote{\pgplot\ --
\url{http://www.astro.caltech.edu/~tjp/pgplot}}, and third-party
commercial packages such as IDL\footnote{IDL --
\url{http://www.rsinc.com/idl}} and
MONGO/SuperMONGO\footnote{SuperMONGO -- \url{http://www.astro.princeton.edu/~rhl/sm}}.

The dominance of the 2d display paradigm is a simple consequence of
both the primary publishing medium -- paper -- and display device --
computer monitor -- being inherently flat and 2-dimensional.
Additionally, the computational demands of 2d graphics display are
typically much smaller than those of three-dimensional (3D) display.
In this paper, by ``3D display'' we mean that a set of geometrical 
objects (or ``geometry'') is described
-- by the programmer and/or software user -- to the underlying
graphics system using a full three-dimensional coordinate system.  The
device itself may then produce a 2d view of the content (e.g. on
a standard desktop monitor); it might render the content to an
immersive display (e.g. a dome) that gives implicit depth perception
cues; or it might produce a genuine {\em stereoscopic}\/ visualization
that presents slightly different views of the geometry to the viewer's
left and right eyes.

Historically, the few 3D visualization tools that were developed did
not achieve widespread use because processing speed limited their
interactivity.  For example, the Karma {\sc xray}\footnote{Karma {\sc xray} --
\url{http://www.atnf.csiro.au/computing/software/karma}} package, 
developed more than ten years ago for volume rendering tasks, was not 
always able to reach usable rendering speeds on the desktop workstations 
available at that time, as it relied on the processing power of early 
Sun SPARC processors.  

This is no longer the case: the astronomers' publishing medium now
enables web-based colour graphics, animation and even interactive
content; cheap 3D display devices are becoming available; and the
computational demands of 3D graphics are now easily handled by
off-the-shelf, hardware-accelerated graphics cards that ship with
nearly every desktop and laptop computer sold today.  The explosion of
availability of very fast graphics cards, fed by the massive computer
game and entertainment market, now means that the most powerful
processor in a desktop computer is often the graphics processor, not
the CPU.  With specialized circuitry for placing graphics primitives
into a 3D virtual space and rendering a view of that space directly to
the screen buffer, displaying 3D environments on desktop computers is
now practical and effective.

Evolution of the academic publishing medium, together with the
wider availability of advanced computer graphics hardware, has
led us and our collaborators to pursue the 3D display,
analysis and publication paradigm.  Specifically:
\begin{itemize}
\item Beeson et al.\ (2003) report on the development
and implementation of a distributed volume rendering technique with
stereoscopic support;
\item Beeson et al.\ (2004) present a web-based tool for presenting
multi-dimensional catalogues in 3D form via VRML technology; and
\item Fluke et al.\ (2006) report on new, economical versions of
traditional display techniques such as dome and tiled wall displays.
\end{itemize}

The Virtual Observatory paradigm (Quinn et al.\ 2004) is
also suggesting new ideas, and VOPlot3D\footnote{VOPlot3D --
\url{http://vo.iucaa.ernet.in/~voi/VOPlot3D_UserGuide_1_0.htm}} is an
interesting development.  It is a Java applet that presents a 2d
projection of a 3D point-based data set (extracted from a
user-supplied multi-column catalogue) and enables rotation, panning
and zooming of the view.  In Rixon et al.\ (2004), the re-casting of
the distributed volume renderer [dvr, Beeson et al.\ (2003)] as a
Grid-enabled application for remote server-based visualization was
presented and discussed.  The Remote Visualization
Service\footnote{RVS -- designed and developed by the CSIRO Australia
Telescope National Facility, see
\url{http://www.atnf.csiro.au/vo/rvs/}} is a nice example of the
adaptation of sophisticated, legacy astronomy visualization software to
the server-side visualization model, and while it does not (presently)
offer 3D viewing capabilities, the plans for RVS always included the
eventual incorporation of volume rendering (via dvr) as an advanced
feature.

A high-profile example of a multi-dimensional data visualization
system is provided by the UK National Cosmology Supercomputer (COSMOS --
\url{http://www.damtp.cam.ac.uk/cosmos/Public/index.html}).  This
system builds on the server-side visualization model to provide access
to multiple visualization services from low-end, graphically
``primitive'' desktop computers.  The user (client) runs one of IRIS
Explorer, Partiview or Vis5d+ on their workstation, and controls the
rendering parameters locally, but the visualization itself is
generated on the server, which provides substantial processor, memory,
disk and graphics resources.  The rendered image is compressed and
transferred back to the client machine for display.  This is a nice
but central-resource-hungry system that provides non-stereoscopic 3D
display services.

The visualization field in astronomy
has become very active recently.  Yet on the whole, the uptake of the
new programs for displaying and analysing data is exceedingly low, and
it is fair to say that many in the community would view them as
``toys'' - they demonstrate a neat or useful idea, and might be
useful occasionally.  They generally do not yet consider any of these
systems as standard, in the league of \pgplot, {\sc miriad},
{\sc IDL} and so on.  Because of this the new tools are not installed
and potential users are not exposed to them.  Consequently  there
is no {\em de facto}\/ standard library for 3D visualization
and display.  As an aside, many of the above tools are actually very
simple to use once installed, and on the whole are very well
documented.

\subsection{Community behaviour}

In view of the above, it was felt that consulting the community to
assess their {\em current}\/ practices in astronomy data visualization
and analysis might assist in the future development of visualization
tools that astronomers will find useful and {\em want}\/ to adopt.  To
this end, Fluke et al.\ (2006) report on their Advanced Image Displays
for Astronomy (AIDA) survey of Australian astronomers.  While based on
a moderately small sample of forty-one astronomers, it is the best
information we have on the behaviour and habits of the community in
this domain.

{\em Visualization and analysis environment.}  Excluding
wavelength-specific software from Fluke et al.\ (2006) Appendix~A
Table~4, we find that the most common analysis tools are: Custom
\pgplot\ tools (regularly used by 44\% of respondents), Karma (39\%),
Mongo/SuperMongo (34\%), IDL (29\%) and other locally developed software 
(27\%).  With the exception of Karma all of these are actually 
programming or scripting environments or libraries that provide
functionality to use from program code. In fact, Karma {\em
does}\/ provide a programming library and interface, but we know of no
current Australian astronomers using the Karma library as opposed to its
pre-packaged tools such as {\sc kvis}, {\sc xray}, etc.  This is a 
stark contrast
to the recent developments described earlier, which were all
pre-packaged, ``out-of-the-box'' applications for displaying or
analysing data,  none of which provide anything like a
well-documented {\em programming}\/ interface or scripting language.
Yet the results of the AIDA survey suggest a preference amongst
astronomers for this approach!

{\em Visualization and analysis plot type.}  From Table~5 of
Fluke et al.\ (2006), we learn that the dominant graph/plot types used
by astronomers are 2d graphs, histograms and plots (used by 95\% of
respondents).  Beyond that, 85\% regularly display/produce 2d images,
and 27\% 3D images.  We expect that the respondents using ``3D
images'' as a visualization method are including both 2d slices 
from 3D datasets, and also volume rendering.  There are no major
surprises here.

{\em Display device.}  
Participants in the AIDA survey were asked about their experience of
display devices, ranging from the ubiquitous CRT/LCD desktop monitor
and single-screen stereoscopic projection, to the sophisticated and
expensive Virtual Room system.  While very few had used anything
beyond the standard issue desktop monitor, $\sim 50$\% of respondents
had seen a single-screen stereoscopic projection system in action but
had not used such a device personally.  Approximately 70\% had seen a
digital dome projection system, as used in planeteriums, but again,
had not used the device for their own work.  Participants were then
asked what factors prevented them from using advanced image displays
in their research.  Lack of knowledge of available displays was the
most common response ($\sim 70$\% of respondents), followed by lack of
appropriate software, lack of local facilities and cost (each nominated by
$\sim 30$\% of respondents).  Comments made by participants
highlighted ``the time required to develop appropriate tools to take
advantage of these displays'' and ``[the] lack of knowledge of
available software'' as key factors preventing higher uptake of
advance display use in astronomy in Australia.

\subsection{Responding to the community}

Based on the outcomes of the AIDA survey, we set about deducing what
capabilities were most needed in a package that would positively
encourage astronomers to begin using advanced, 3D plotting techniques.
With recent experience of developing several advanced visualization
tools, and the skills and knowledge in advanced displays built up at
our institute, we have set about sharing this capability with the
astronomy community in a form that we hope will experience significant
uptake.  That is, we have created an advanced visualization library 
with the following key features and benefits:
\begin{enumerate}
\item{\em provides three-dimensional plotting functions} --- dramatically
increases capacity and options for display of multi-dimensional data.
\item{\em provides a mechanism for navigating the 3D world} ---
drastically improves comprehension of 3D geometry, and perception of
depth on 2d devices.
\item{\em is a programming library} --- can be called from existing
and new C, FORTRAN and C++ codes.
\item{\em has a PGPLOT-like interface} --- many of the functions will
look familiar to PGPLOT users.
\item{\em works on GNU/Linux and Apple/OSX} --- available for the most
common astronomer workstations/laptops.
\item{\em works on standard monitors} --- enables users to create and
explore 3D geometry at their desktop, without requiring expensive
hardware.
\item{\em works on advanced displays} --- if a stereoscopic or dome display is
available, it can be used with no change to the code.
\item{\em comes with documented sample programs} --- allowing new users to
learn by example.  
\item{\em is a binary distribution} --- no compilation is necessary to
install the library or run the sample programs.
\item{\em can save hardcopy output} --- for production of figures and movies 
for journals and websites.  The TGA format is used.
\item{\em can save geometry files} --- allowing views to be saved to
disk and displayed later without re-running code.
\end{enumerate}

VTK\footnote{VTK -- \url{http://public.kitware.com/VTK}} offers a
number of the above features, but it is an extremely heavyweight
solution, with an extended learning curve.  Additionally, VTK is
written in C++, which is not the most popular language of the astronomy
community.  IDL (versions 5 and above) now has support for 3D
graphics, and {\em in principle}\/ with some clever coding can
generate stereoscopic pairs for projection.  We have not seen IDL
produce stereographics though, and would be interested to hear from
astronomers who have done so.

Tipsy\footnote{Tipsy --
\url{http://www-hpcc.astro.washington.edu/tools/tipsy/tipsy.html}} is
an application specific to N-body simulations: it can display particle
positions and velocity vectors from an arbitrary viewpoint.  The user
can follow selected particle(s) as they evolve, and calculate bulk
properties of particle groups.  This specialized tool is well-used
throughout the N-body community, but cannot be programmed to do new
things.  It would be moderately simple to reproduce the capabilities
of Tipsy using the \splot\ library, thus gaining support for dome and
stereoscopic displays, and the capability for extensions to the
application.

This paper briefly describes the architecture and implementation of
\splot\ --- a library that meets the above specifications and is being
made available (in binary form only) to the astronomical community.  
We then give some
specific usage examples, consider some future extensions and uses, and
close the paper with directions on obtaining the \splot\ distribution.

\section{Design and Architecture}

\subsection{STEREO2}

For the last seven years the Centre for Astrophysics and
Supercomputing at Swinburne University of Technology has been
responsible for providing visualization services to its own
researchers and to the wider Swinburne research community.  From time
to time, it has also delivered visualization services to external
users, including the Anglo-Australian Observatory, and to external
researchers/collaborations, including Brent Tully and the 6dF team.
Consequently, Swinburne has built up a small but substantially capable
set of applications for displaying data from many different sources.
Almost without exception these visualization solutions support one or more of
Swinburne's advanced display devices (viz.\ passive and active
stereoscopic projection displays, MirrorDome, and The Virtual Room),
as well as working on normal desktop workstation displays.

\begin{figure}[ht]
\begin{center}
\includegraphics[scale=0.6, angle=0]{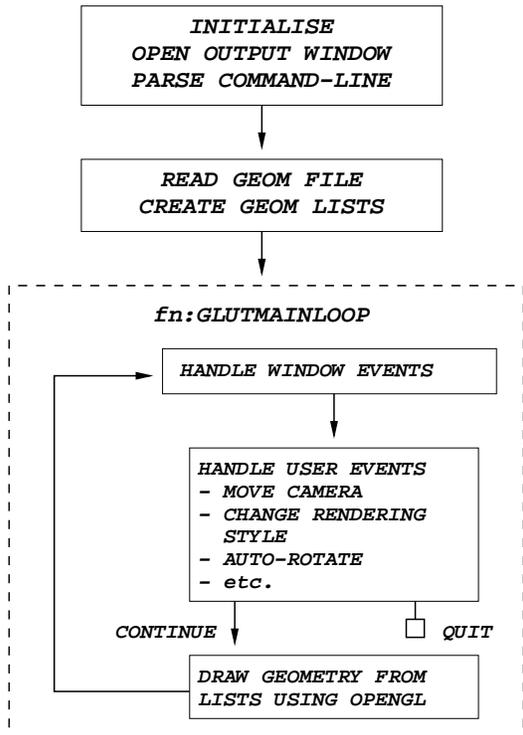}
\caption{Block diagram of the \stereotwo\ program flow.}\label{fig:stereo2}
\end{center}
\end{figure}

One of these tools -- \stereotwo\ -- has a simple but effective
architecture that seemed amenable to conversion to a functional
programming library.  A simple block diagram of the \stereotwo\
control flow is shown in Figure~\ref{fig:stereo2}.  Very briefly, 
\stereotwo\ reads a disk file containing a description of the geometry
to be rendered, stores the geometry in a set of lists of different
geometry types (e.g. ball, line, quad, textured quad), and then enters
its main loop.  The main loop draws the geometry using
OpenGL\footnote{\url{www.opengl.org}} calls, handles any events coming
in via the keyboard or mouse, and loops to draw the geometry again,
taking into account any changes (e.g. to the camera location or
orientation) made in response to handled events.  The main loop is
terminated by pressing a special key (`{\tt q}' or `{\tt ESC}') in the
display window.  \stereotwo\ supports twenty-one 3D graphics primitives
from input referred to as GEOM files.

\stereotwo\ is fast, fully-featured in terms of graphics
primitives, and supports 2d and stereoscopic display devices.  It has
been built on GNU/Linux, Apple/OSX and Compaq/Tru64 platforms, and has
been used regularly at Swinburne for more then three years.  It is
well-tested and robust, and has been deployed in binary form as a
component of scientific visualization theatres installed at sites
around Australia and overseas.  \stereotwo\ is at ``end of life'' and
so its code-base is stable and unchanging except for occasional minor
bug fixes.

\subsection{S2PLOT}

\begin{figure}[ht]
\begin{center}
\includegraphics[scale=0.6, angle=0]{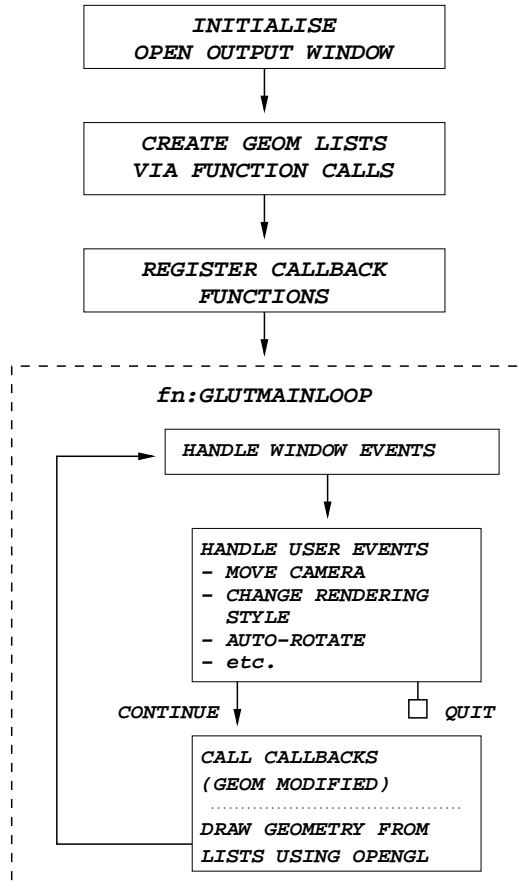}
\caption{Block diagram of the \splot\ program flow. }\label{fig:s2plot}
\end{center}
\end{figure}

The \stereotwo\ architecture is easily adapted to provide the core of
a three-dimensional plotting library.  In some sense it is a
trivial change: replace the GEOM-format file-reading code with a
functional interface whose components explicitly add elements to the
lists of geometry types.  A simple block diagram of the resulting 
\splot\ control flow is shown in Figure~\ref{fig:s2plot}.

However, there are choices in the style of functional interface.  The
simplest approach was to write a set of functions that map directly to
the 21 primitive geometry types of the GEOM format.  We felt
that only doing this would miss the opportunity to present an
interface that would immediately look familiar and even ``friendly''
to many astronomers.  Instead we designed a functional interface
that replicated many of the functions provided by \pgplot, but using a
3D coordinate system.
The viewport ({\tt pgsvp}) and world-coordinate ({\tt pgswin}) functions
of \pgplot\ have obvious 3D generalisations in \splot.
Consider the \pgplot\ function 
\begin{quote}
\tt pgline(n,xpts,ypts)
\end{quote}
that draws {\tt n} connected line segments, joining the points {\tt
(xpts[i],ypts[i])} and {\tt (xpts[i+1],ypts[i+1])} for {\tt i = [1,
n-1]}.  It suggests the \splot\ function
\begin{quote}
{\tt s2line(n,xpts,ypts,zpts)},
\end{quote} 
that draws {\tt n} connected line segments in a 3D space.  Similarly,
the \pgplot\ function
\begin{quote}
\tt pgrect(x1,x2,y1,y2)
\end{quote} draws a rectangle extending from {\tt (x1,y1)} to
{\tt (x2,y2)}.  It suggests the \splot\ function 
\begin{quote}
\tt s2rectxy(x1,x2,y1,y2,z)
\end{quote} that draws a rectangle in the {\tt xy}
plane, extending from {\tt (x1,y1,z)} to {\tt (x2,y2,z)}.  Similarly
it suggests \splot\ functions 
\begin{quote}
{\tt s2rectxz(x1,x2,z1,z2,y)}, and \\
\tt s2rectyz(y1,y2,z1,z2,x) 
\end{quote} 
for drawing rectangles in the {\tt xz} and {\tt yz} planes.  
The obvious extension to a more general polygon in 3D is
made available through a ``native'' routine (i.e. one that provides a
direct analog of a GEOM-format primitive)
\begin{quote}
{\tt ns2Vf4(XYZ *P, COLOUR col)}, and \\
{\tt ns2Vf4n(XYZ *P, XYZ *N, COLOUR col)}
\end{quote}
that draws a coloured four-vertex facet with normals calculated either
automatically ({\tt ns2Vf4}) or specified by the user ({\tt ns2Vf4n}).

With this design, we can go well beyond providing a set of 3D graphics
primitives.  The \splot\ library provides a world with
(linear) coordinates of the user's choice, and high-level functions
that produce parametric line and surface plots, sampled surface plots,
axis labels, error bars, colour wedges and so on.  The complete set of
\splot\ functions based on \pgplot, together with the native
functions, provide a comprehensive set of functions for generating 3D
realisations of data sets.  Not {\em all}\/ of the \pgplot\ primitives
were suitable for the 3D implementation: some lack useful counterparts
in 3D visualization (e.g.\ PGBIN, PGHIST), some are device-specific and
not well-matched to (or even possible with) the output device for 3D
graphics (e.g.\ PGSCLP), and others are concerned with interactive
input mechanisms that have deliberately been left out of this first
release of \splot\ (e.g.\ PGCURS).

\section{Implementation and Use}

{\em Output device.}  \splot\ uses OpenGL calls to display 3D
geometry, and compiles on GNU/Linux and Apple/OSX systems.  On Linux,
its only direct output device is an X-Windows display with the {\tt
GLX} extension.  This extension is usually provided by the Mesa 3D
Graphics Library\footnote{\url{http://www.mesa3d.org}} for software
mode (unaccelerated) OpenGL, or by drivers specific to the installed
graphics card that provide hardware mode (accelerated) OpenGL calls.
On a standard OSX installation, the direct output device is the Aqua
screen.  By virtue of its \stereotwo\ heritage, \splot\ provides the
following ``virtual'' output devices:
\begin{itemize}
\item {\tt /S2MONO} -- non-stereo, perspective display;
\item {\tt /S2ACTIV} -- active stereo, perspective display;
\item {\tt /S2PASSV} -- passive stereo, perspective display;
\item {\tt /S2FISH} -- non-stereo, full fisheye projection;
\item {\tt /S2TRUNCB} -- non-stereo, truncated-base fisheye
projection; and
\item {\tt /S2TRUNCT} -- non-stereo, truncated-top fisheye
projection.
\end{itemize}
Non-stereo display is always possible and is suitable for desktop
monitors and single projector audio-visual systems.  Passive stereo mode
produces a double-width window containing left-eye and right-eye views
of the geometry, suitable for display on X-Windows systems that use
extensions (e.g. ``Xinerama'') to deliver output to a crossed
polarisation, two-projector system.  Active stereo refers to
frame-sequential stereo, which is available on desktop and projection
systems with synchronized LCD shutter glasses.  

The {\tt /S2FISH} device is available for use in a system providing
direct projection through an ideal, 180-degree fisheye lens.  The
truncated-base fisheye projection is the projection used in
VisionStations, and the truncated-top projection is that used by
planetariums with fisheye lenses.  Selectively truncating the fisheye
projection enables some optimisation of resolution across the
projection surface at the cost of typically one-quarter of the display
surface remaining unused.  Swinburne's dome projection device,
MirrorDome, {\em is}\/ supported by \splot, but only at licensed
sites.

Full-screen display mode is available with all devices by appending
the character {\tt F} to the device name.  It is to some extent
dependent on the window manager in use.  With appropriate hardware,
the ``flat screen'' display devices ({\tt /S2MONO*}, {\tt /S2ACTIV*}
and {\tt /S2PASSV*}) can be used to drive multi-projector systems
whose individual outputs are ``tiled'' to generate a large,
very-high-resolution display.  For example, an off-the-shelf Apple G5
system equipped with an XGA splitter on each of its two video outputs
can drive four XGA projectors to deliver a display of $2048 \times
1536$ pixels.  By adding three dual-output video cards to the G5's
expansion slots, this display can be extended to $4096 \times 3072$
pixels!  Most of the non-full-screen modes are made available only for
testing and development.  Almost all production projection systems use
full-screen mode to obtain maximum resolution across the projection
surface.

Hardcopy output is available, either in ``one-shot'' mode or
continuous recording.  The displayed window (including both eye views
for stereo mode rendering) is saved to an image file having the
same dimensions as the display window.  Geometry can be saved to a
GEOM-format file for later viewing with the included utility {\tt
s2view}, and camera location and configuration can also be saved.
Continuously-recorded images can later be assembled into
movies using third party utilities.

{\em Device and world coordinates.}  The primary \splot\ state
variables include the location of the current viewport (the part of
the 3D world that is being drawn to, defined by the two corners of a
3D box, set by {\tt s2svp}) and the world coordinate values that
correspond to the corners of the viewport (set by {\tt s2swin}).
All \splot\ functions that produce geometry are called with world
coordinates that are internally and linearly transformed to
``device'' coordinates using the stored state.  Since we are concerned
with drawing a 3D environment on a 2d screen, there is no real concept
of a ``pixel'' in the 3D space, so device coordinates need not be
rounded integers.  In fact, the device coordinate system as set by
{\tt s2svp}, is largely of no concern to the \splot\ user unless
they need to display geometry in a non-cubic space, or they wish to
place more than one plot in the 3D space.  Once the geometry is
complete and \splot\ is asked to display the geometry (see below) the
camera location and field-of-view will be set to automatically show
the part of the world that is populated with geometry.

The user is free to set a world coordinate system that produces
non-cubic environments.  In such coordinate systems, the specification
of the radius of disks, cones and spheres is not well-defined.
\splot\ makes a ``best effort'' by calculating the quadratic mean of
the radius converted to device coordinates on each of the concerned
axes.  For example, the radius of a disk in the {\tt x-y} plane in
device coordinates is the quadratic mean of the requested (``world'')
radius converted to device coordinates on the {\tt x} and {\tt y}
axes.  Similarly, the radius of a sphere in device coordinates will be
the quadratic mean of the requested radius converted to device
coordinates on all three axes.  Some functions (e.g. {\tt s2circxy})
provide control over this calculation though, and enable the user to
draw an ellipse in world coordinates that may or may not look like a
circle when viewed (in device coordinates), and to control the aspect
ratio of this ellipse.  In general, the user will make life much
easier for themselves by calling {\tt s2svp} and {\tt s2swin} with
equal length axes!

{\em Colour and light.}  OpenGL provides true 24-bit colour.
Thus, any colour requested will be available, and colourmaps (e.g. for
surface plots) can have thousands of entries to produce smooth colour
gradients.  Presently, an arbitrary limit for colourmap size is set at
16384 colours.  The lighting for the 3D geometry comprises ambient
light of a given colour and intensity, and up to eight point-source
lights also of given colour placed in the world.

{\em Labelling.}  One of the draw-cards of the \pgplot\ library
is its capability to label graphs clearly and sanely.  We have
included a number of functions in \splot\ that are inspired by
\pgplot\ counterparts, including {\tt s2box} to draw and label the
viewport with world coordinates, {\tt s2iden} to write the username
and creation date of the plot on the ``page'', {\tt s2lab} to draw a
title at the top of the ``page'', and {\tt s2env} that provides a
convenient wrapper for {\tt s2swin} and {\tt s2box}.

{\em Callback functions.}  There is one major difference between
the \splot\ and \pgplot\ programming models.  \pgplot\ allows the user
to draw some geometry and show it to the user, and then continue to
add geometry to the plot.  Conversely \splot, via the \stereotwo\
code-base, uses the main event loop function ({\tt glutMainLoop}) of
the GLUT library\footnote{OpenGL Utility Toolkit, originally written
by Mark Kilgard of SGI}.  This function is responsible for processing
and dispatching events to registered handler functions, including
refresh and redraw events, and mouse and keyboard events that may
modify the camera or geometry settings.  The {\tt glutMainLoop}
function {\em never returns}, and as a consequence can be called only
once per program invocation.  Furthermore, the 3D geometry stored in
\splot's internal lists are not displayed until this loop is entered.
Hence the programming model, as it stands, requires the user of the
\splot\ library to place {\em all}\/ of their geometry creation
function calls {\em prior}\/ to their call to the \splot\ function
{\tt s2show}, which implicitly calls {\tt glutMainLoop}.  \pgplot\
programmers would be quite used to a less constrained model where
graphics can be flushed to the device at any time, and as many times
as desired.  The
\pgplot\ concept of erasing the device or moving to a new page is not
present in \splot.

To remove this limitation of \splot, we have implemented an elegant and
powerful mechanism to hand some control back to the programmer.  The
\splot\ {\em callback system}\/ allows the user to register their own
function to be called once per refresh cycle. If there is no user
intervention or excessive CPU load, this refresh occurs up to 25 times 
per second.  The
callback function is called by the \splot\ library prior to redrawing
the geometry, and two arguments are given to the function: the current
time (in seconds) and the number of times a special key (currently the
space bar) has been pressed.  The programmer can use any
\splot\ drawing commands within the callback to produce new geometry
for display.  The time value can be used to calculate new positions for
geometry, and the number of key presses to change some arbitrary state
(if desired).  We differentiate between {\bf static}\/ geometry that is
{\em created}\/ prior to the one and only call of the {\tt
s2show} function but {\em displayed}\/ every refresh cycle, and
{\bf dynamic}\/ geometry that is recreated {\em and}\/ drawn each and
every refresh cycle.

Three key features of the callback function make this a powerful
approach to producing scientific and educational 3D content:
\begin{enumerate}
\item the callback function itself can register another callback function in
its place, thus enabling a flow of control from one callback function
to another;
\item the callback function can disable the callback mechanism,
effectively freezing the dynamic geometry until the user intervenes
and re-enables the callback by hitting a special key (currently `z');
and
\item the callback function can use static local variables in C, or
{\tt COMMON SAVE}'d variables in FORTRAN, to preserve state and data from one
call to the next.
\end{enumerate}
Used creatively, these three capabilities can produce innovative
programs for display and presentation of data.  Some particular ideas
will be discussed in Section~\ref{sec:eg}.

{\em Linear programs.}  The callback structure can be confusing
at first, and most astronomers will not be used to planning or writing
callback code.  Even though the \splot\ callback system is
straightforward, control of execution is never really available and a
linear program structure is still desirable in many cases.
Fortunately, two common implementations of the GLUT library provide an
extension that make linear programming possible.  {\em Freeglut},
which is standard with many GNU/Linux systems, provides a new function {\tt
glut\-Main\-Loop\-Event()} that ``processes a single iteration of the event
loop and allows the application to use a different event loop
controller or to contain re-entrant code''.\footnote{{\em freeglut} API
reference, \url{http://freeglut.sourceforge.net}}  Similarly, Apple's
implementation of GLUT for Apple/OSX includes a new function {\tt
glut\-Check\-Loop()} that provides precisely the same behaviour.
Hence for \splot\ installations on Apple/OSX and on GNU/Linux
with {\em freeglut}, the following code fragment
becomes possible, and thus the more standard
\pgplot\ programming model is provided:

{ \footnotesize
\begin{verbatim}
s2open(...); /* open s2plot device                 */
s2swin(...); /* set world coordinate range         */
...          /* various calls to create geometry   */
s2disp(...); /* display the geometry, let the user */
             /* interact, and return control when  */
             /* the TAB key is pressed or a time-  */
             /* out occurs                         */
s2eras(...); /* erase existing geometry            */
...          /* add further geometry to the world  */
s2show(...); /* display again, and this time       */
             /* relinquish control to GLUT         */
\end{verbatim}
}

Since the {\tt s2disp} function uses non-standard extensions to GLUT,
it should be considered an advanced (expert) feature in \splot\
and programmers are encouraged to use the callback model where 
possible.

\begin{figure}[ht]
\begin{center}
\includegraphics[scale=0.40, angle=0]{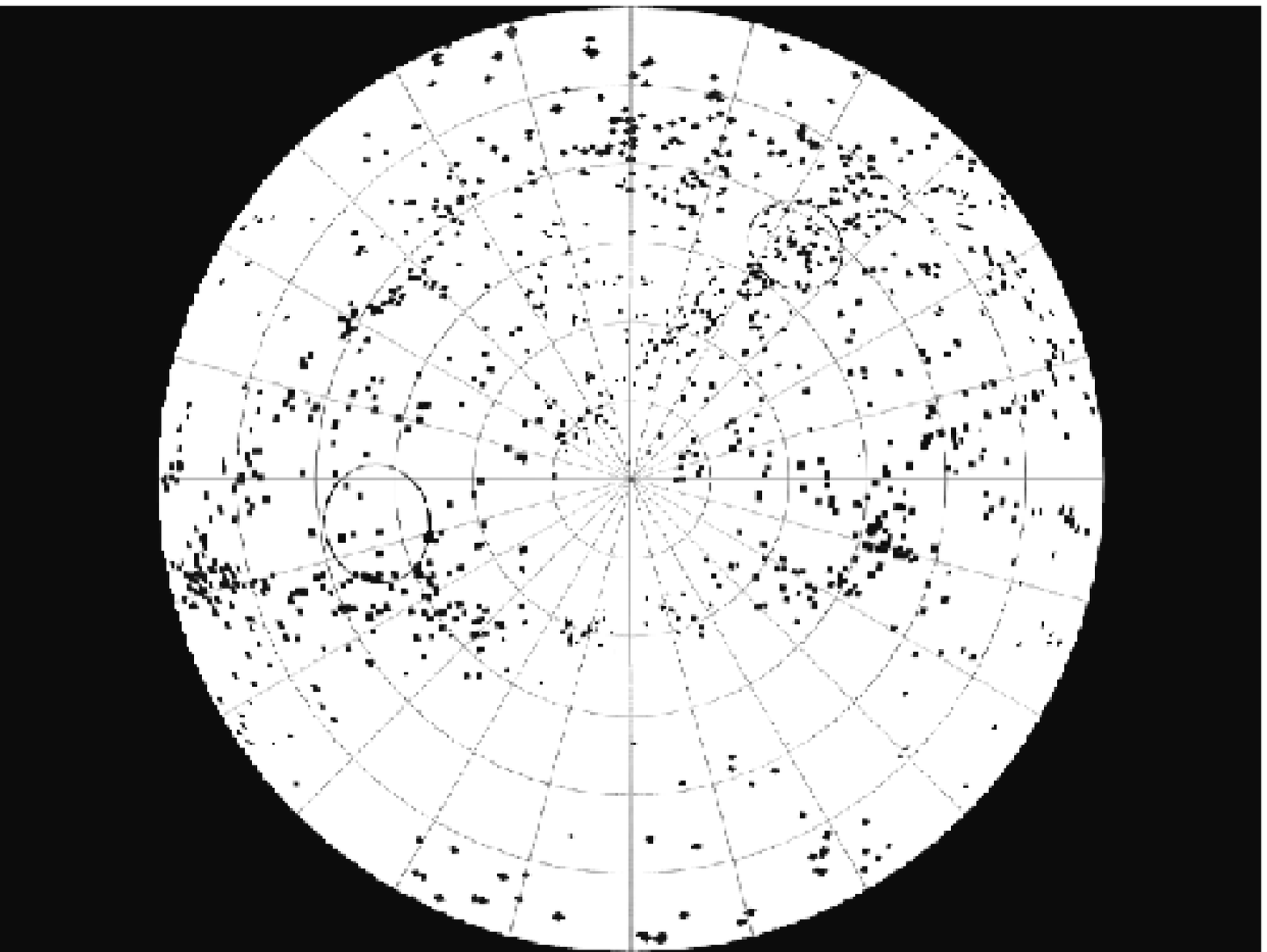}

\includegraphics[scale=0.40, angle=0]{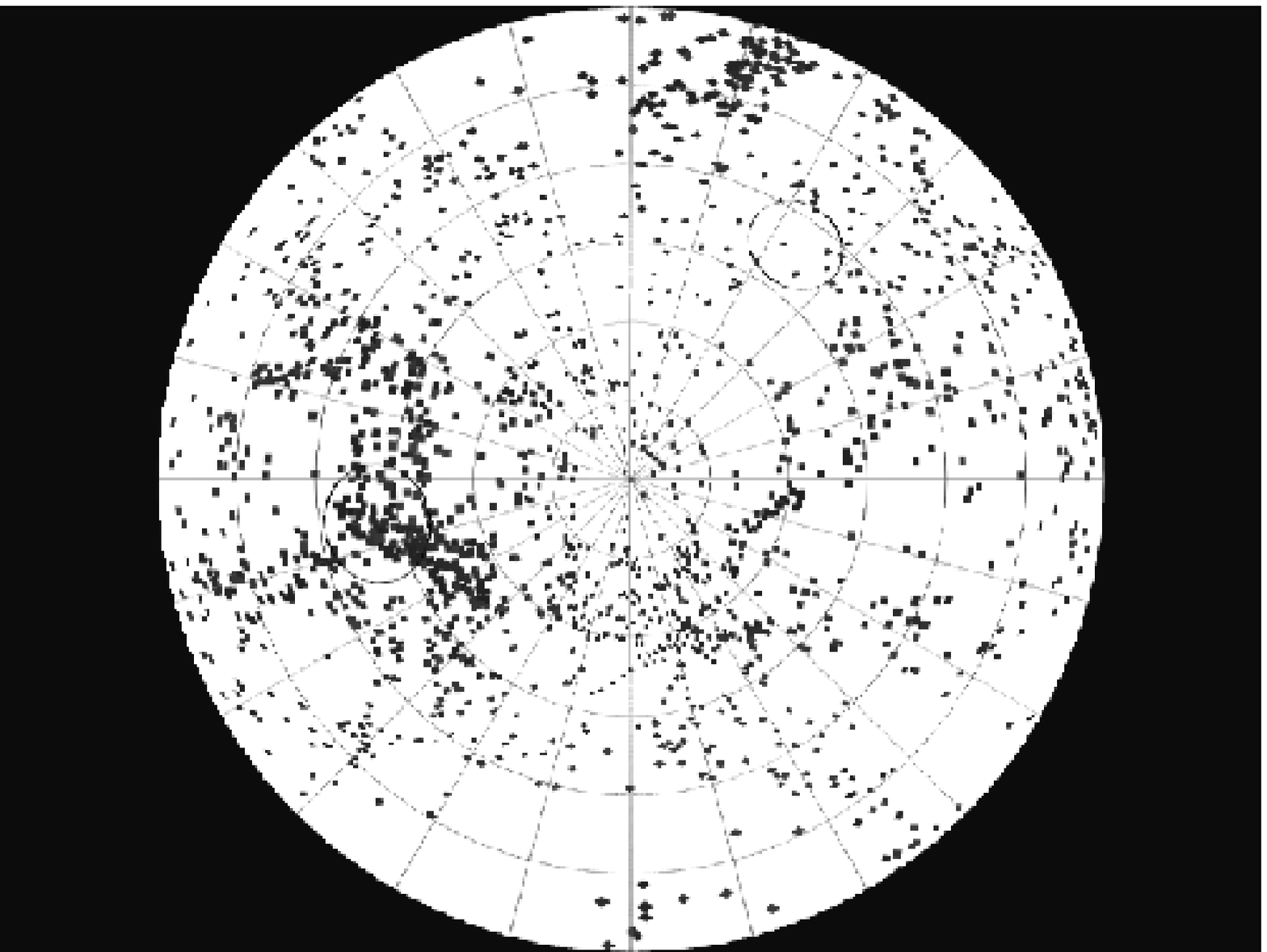}
\caption{Fisheye projections of southern HICAT galaxies.
(Top) Velocity range 0--2000~km~s$^{-1}$.  The Fornax cluster is 
visible in the top-right circle. (Bottom) Velocity range
3000--6000~km~s$^{-1}$.  In this velocity range, the Fornax cluster is
no longer evident and the Centaurus cluster now dominates the sky
(centred in the circle to the left of centre).\label{fig:hicat-for}}
\end{center}
\end{figure}

\begin{figure}
\begin{center}
\includegraphics[scale=0.40, angle=0]{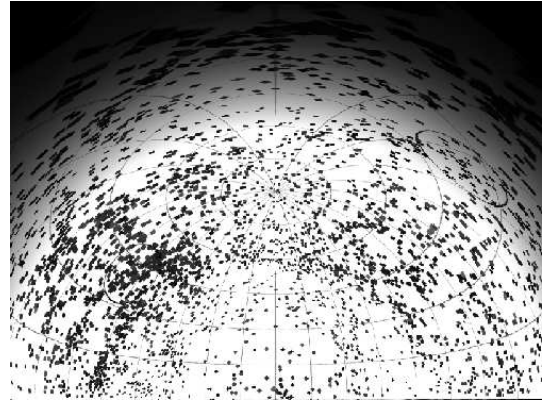}
\caption{Projection of southern HICAT galaxies
in the velocity range 0--10000~km~s$^{-1}$ suitable for display 
with a Swinburne MirrorDome.  Note the extensive (but correct) distortion
towards the top and sides of the image where glancing projections off
the mirror occur, and the deliberate dimming of the image to maintain
uniform brightness over the dome.\label{fig:hicat-warped}}
\end{center}
\end{figure}

\section{Examples}
\label{sec:eg}

The possibilities for using the \splot\ in astronomy and science more
generally are essentially limitless.  \splot\ can be used for almost
anything from a simple, static plot of the spatial distribution of
objects found in a survey such as 2dF, to a dynamic, interactive
conference presentation that combines traditional ``powerpoint slide''
content with pictorial animations running alongside real-time slices
of an observational data set.  It would even be quite straightforward
to integrate \splot\ visualization code with data acquisition systems
to provide, for example, real-time and multi-dimensional
representations of instrument status, data quality, or survey
progress.  Native support for economical stereoscopic devices and dome
projection systems, {\em needing no additional programming}, opens new
horizons in data exploration and visualization for individual
scientists and teams of scientists, while ``save now and view later''
features mean visualizations can be shared around
geographically-spread collaborations.  

We now describe some examples of \splot\ use, gratefully
received from current users at our institute.  Code for some of these
examples is available in the \splot\ distribution.
Where suitable, example output is illustrated in this paper in figure form.  
But we remind readers that static views of three-dimensional geometry are 
usually a poor substitute for having the ability to view and rotate 
the data directly on-screen.

\subsection{Dome display of an all-sky survey}

The HI Parkes All Sky Survey Catalogue (HICAT; Meyer et al.\ 2004) is
a recent galaxy redshift survey that lends itself well to advanced
display and exploration methods.  Since it is a survey of the entire
southern sky, we have chosen to use it as an example of the dome
projection capabilities of the \splot\ library.  Like any wide-field
redshift survey, HICAT can be used to visualize large-scale structure
at different redshifts.  We show in
Figures~\ref{fig:hicat-for} and \ref{fig:hicat-warped} sample output of a
simple \splot\ program that loads the catalogue, plots the galaxies on
the surface of a sphere (the ``sky'') and enables the user to select
that redshift ranges are plotted using the numeric keys.  Pressing
the `{\tt 1}' key toggles the display of galaxies in the velocity
range 0--1000~km~s$^{-1}$, the `{\tt 2}' key toggles display of
galaxies in the range 1000--2000~km~s$^{-1}$, and so on.  The location
of two dominant southern clusters --- Fornax and Centaurus --- are
marked by circles in the figures.    The sample code for this demonstration
is available with the \splot\ distribution ({\tt astro1.c}) -- see
section \ref{sec:distribution}.

\begin{figure}
\begin{center}
\includegraphics[scale=0.4, angle=0]{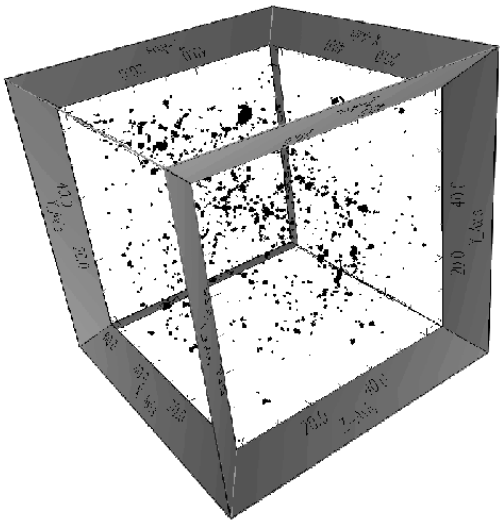}
\caption{Overall view of halo distribution in a $\Lambda$WDM 
simulation in a 64~$h^{-1}$~Mpc box.\label{fig:ar0002i}}
\end{center}
\end{figure}

\begin{figure}
\begin{center}
\includegraphics[scale=0.4, angle=0]{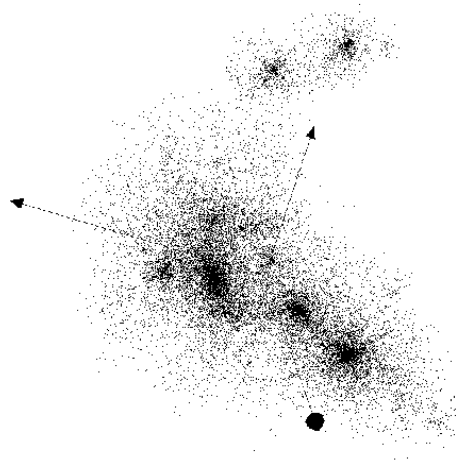}
\caption{Zoomed view of a section of the halo distribution of the
previous figure.  Real-time rotation of the visualization, and/or
stereoscopic display of same, conveys substantially more information
than this figure can.\label{fig:ar0001i}}
\end{center}
\end{figure}

\subsection{The structure of dark matter halos}
\begin{figure*}[t]
\begin{center}
\includegraphics[scale=0.75, angle=0]{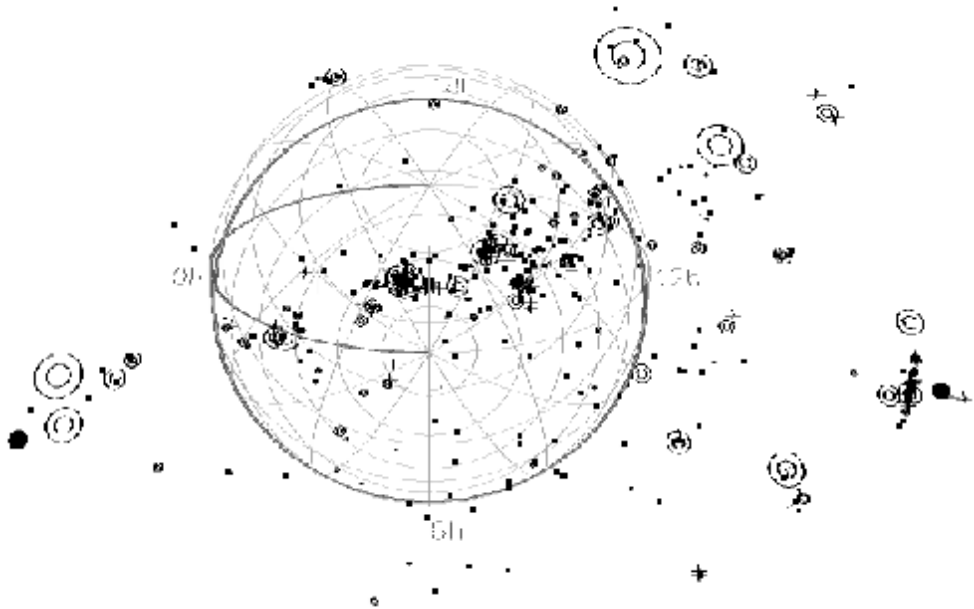}
\caption{Three-dimensional view of the LVHIS targets.  See text 
for a description of the symbols.\label{figA}}
\end{center}
\end{figure*}

\begin{figure*}[t]
\begin{center} 
\includegraphics[scale=0.75, angle=0]{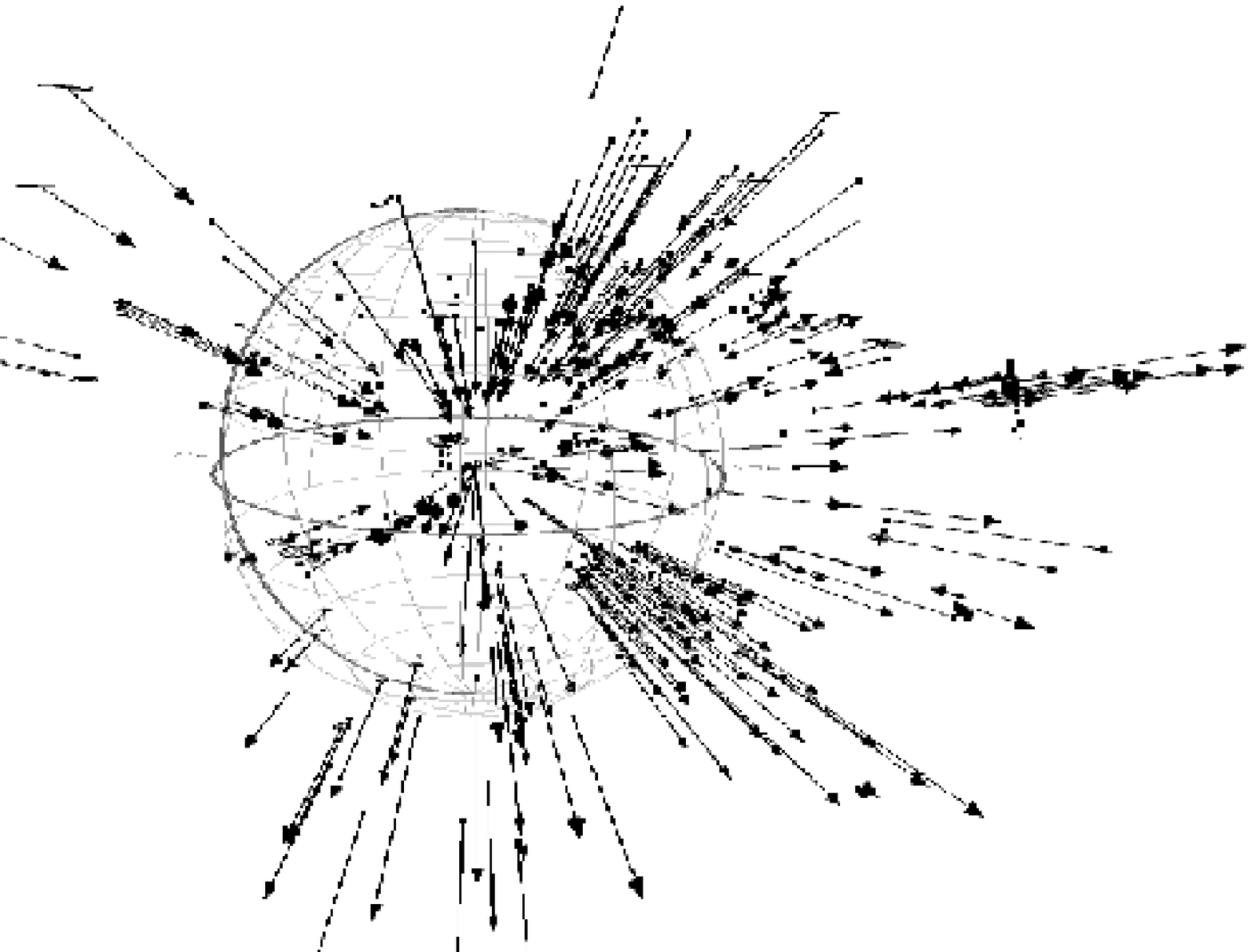}
\caption{Three-dimensional comparison of galaxy distance measures 
for the LVHIS sample.  See text for further information.\label{figB}}
\end{center}
\end{figure*}

Dark matter halos are the three-dimensional structures in which galaxy
clusters and galaxy groups dwell.  Their presence and properties
directly determine the observed structure in the Universe, and thus
the numerical simulation of halo formation can lead to important
constraints on cosmological theories.  The raw and processed output of
N-body simulations is naturally spatial in nature, and is well-suited
to the \splot\ 3D visualization environment.

Ashley Rowlands --- a summer student at Swinburne 2005/06 --- has
compared the shapes of dark matter halos in $\Lambda$CDM and
$\Lambda$WDM cosmologies.  A friends-of-friends algorithm was used to
generate catalogues of dark matter halo masses and positions from the
zero-redshift simulation output, and \splot\ was then used to display
the resulting physical halo distributions.  For both the new student
and his ``seasoned'' supervisors, being able to explore the structures
so quickly and easily, directly from their program code, was
reportedly very useful.  Sample \splot\ visualizations of the data are
shown in Figures~\ref{fig:ar0002i} and \ref{fig:ar0001i}.  

\subsection{LVHIS - local volume HI survey analysis}

The Local Volume HI Survey (LVHIS) aims to study gas-rich galaxies in
the local volume ($D \sim 10$ Mpc or radial velocity $V_{\rm LG} <
550$ km s$^{-1}$), through 20-cm continuum and HI line
observations. One source of LVHIS targets is the Catalog of
Neighbouring Galaxies (CNG) by Karachentsev et al.\ (2004), an all-sky
catalog of 451 galaxies with $D < 11.4$ Mpc.  Examples of S2PLOT
visualizations of the CNG dataset are shown in
Figures~\ref{figA} and \ref{figB}.

Figure~\ref{figA} shows the three-dimensional spatial locations of
galaxies using RA, Dec and distance converted into (x,y,z) triples.
Distances are determined from one of: Cepheid variables, luminosity of
the tip of the red giant branch, surface brightness fluctuations,
luminosity of the brightest stars, the Tully-Fisher relation,
membership in known groups or a direct application of the Hubble
relation.  Elliptical galaxies are shown as shaded spheres, S0
galaxies as 3D crosses and spirals with measured angular diameter as
disks (a second disk is present for those systems where an HI flux has
been measured, however, the diameter of these HI disks are not well
known at present), spirals without measured diameters and all other
object types are indicated by points.  The reference grid of right
ascension and declination coordinates has radius 5~Mpc, with RA=0h and
Dec=$0^\circ$ indicated by a thicker line width.  The view is down the
axis from positive to negative declinations.  A deficit of galaxies
between RA 15h and 24h is readily apparent, and the Super-galactic
plane is visible (particularly when viewed interactively or with a
stereoscopic display).  Limitations in the distance model are also
apparent: note the galaxy group in the lower right-hand quadrant where
all group members have been placed at the same distance.

Figure~\ref{figB} shows the differences between the original
measurement of distance, and a value based on the measured
heliocentric radial velocity (converted to distance using Hubble
constant of 72~km~s$^{-1}$~Mpc$^{-1}$ without further velocity
corrections).  The view is taken almost parallel to the equatorial
plane.  Systematic differences between the velocity distance and the
other distance methods are apparent.  The galaxy group identified in
Figure~\ref{figA} is now in the upper right quadrant, and it is
apparent that each group member has a different radial velocity.
There is also evidence of coherent large-scale velocity flows towards
the centre at positive declinations and away from the sphere at
negative declinations.  A future investigation of this dataset could
include an examination of the systemic differences between each
distance method.

\subsection{RAVEing in the ChromaDome}

\subsubsection{RAVE}

\begin{figure*}[ht]
\begin{center}
\includegraphics[scale=0.75, angle=0]{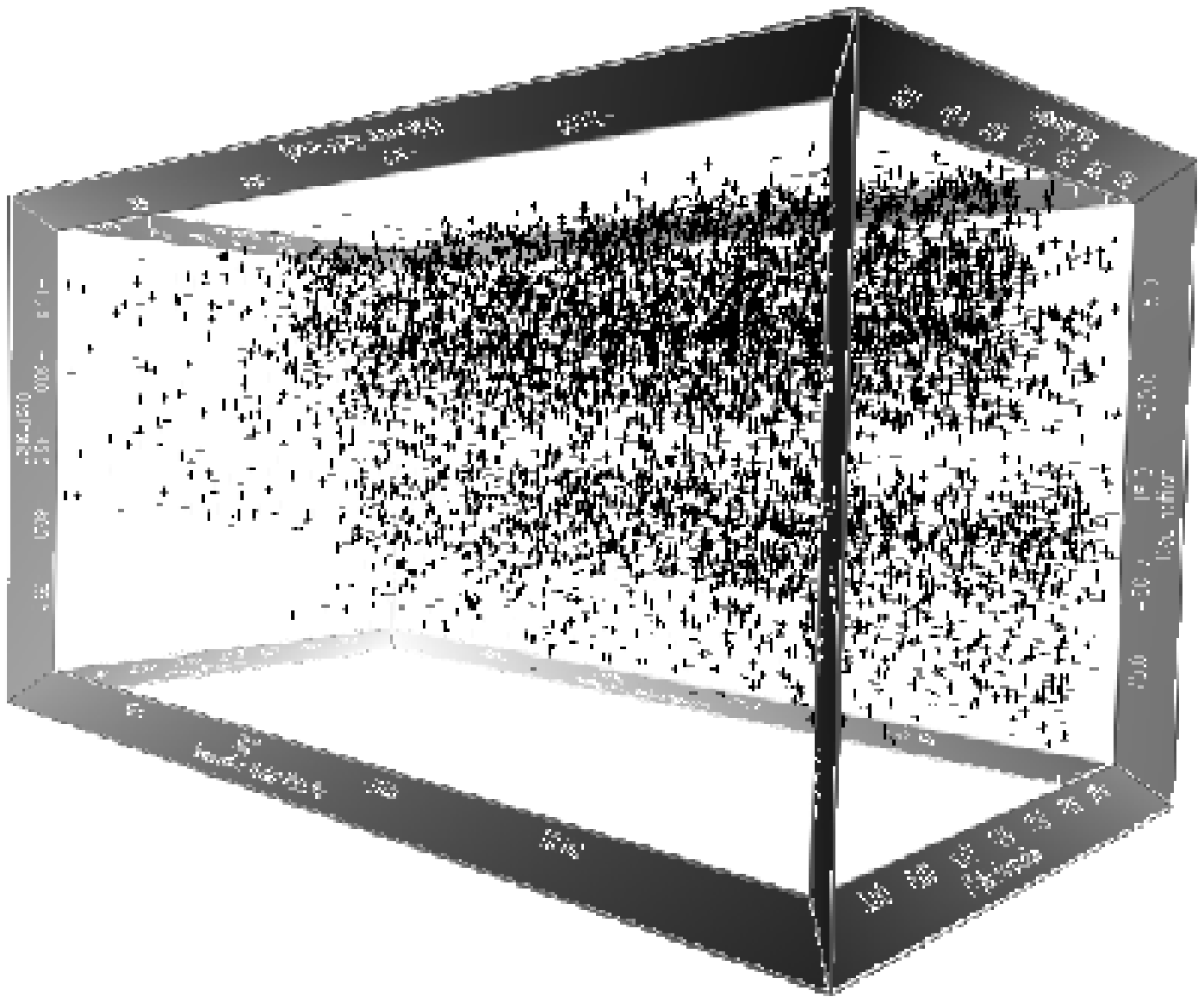}
\caption{The RAVE Data Release 1 (DR1) with radial velocities ($v_{\rm r}$) 
plotted versus celestial coordinates (right ascension and declination).
Only stars with $-50$ km s$^{-1} \leq v_{\rm r} \leq 0$ km s$^{-1}$ are shown. 
\label{figC}}
\end{center}
\end{figure*}

RAVE, the RAdial Velocity Experiment, conducted with the 6dF
spectrograph on the UK-Schmidt telescope in Australia, will provide an
all-sky medium resolution map of the Galaxy.  As an exercise in
determining how straightforward \splot\ is for studying new datasets,
the RAVE first data release (DR1) comprising 25,274 radial velocities
from a region $\sim 4,670$ square degrees was obtained in the week
following its announcement.  After minimal coding, the 
\splot\ visualization of DR1 shown in Figure~\ref{figC} was obtained. 
In this figure,  DR1 stellar positions are plotted in celestial coordinates 
(right ascension and declination) with heliocentric radial velocity,
$v_{\rm r}$, providing the third coordinate. Only stars with  
$-50$ km s$^{-1} \leq v_{\rm r} \leq 0$ km s$^{-1}$ are shown, 
and the dataset is best viewed interactively. 

\subsubsection{ChromaDepth 3-D}

Swinburne has recently been developing economical, multi-person
advanced display devices.  One of these --- the MirrorDome --- is
substantially immersive (offering a $\sim 2 \pi$ steradian view) and it
is desirable to enhance the effect by providing a stereoscopic
display.  However, using traditional ``image-pair'' stereoscopy is
troublesome in the dome, because maintaining a single uniform audience
horizon (i.e.\ eye-separation vector) is essentially impossible.  Even
for a {\em single}\/ observer, image-pair stereoscopy is awkward
because the person can move throughout the dome, and orient their
optical axis arbitrarily.  The required warp for correct dome display
ultimately depends on observer location within the dome together with
roll, tilt and yaw.  Every single refresh cycle of the display would
entail calculating a new warp map to do things properly.  We have not
yet optimised the warp map code to the extent that this is possible in
realtime.

An alternative to image-pair stereoscopy exists in ``ChromaDepth
3-D'', a technique patented by Richard Steenblik in 1983 (US Patent
No.\ 4-397-634).  It amplifies the natural chromostereoscopic effect
that arises from the variation in refraction of different wavelength
light: red light focuses further back than blue light does after
entering a refraction-based optical lens system such as the human eye.
ChromaDepth 3-D is not stereo-pair based: a single image can be
produced that looks sensible without glasses, and stereoscopic with
glasses.  Red objects appear closer (i.e.\ are in the foreground), and
blue objects appear further away (in the background).  Objects having
colours in the rainbow lying between red and blue fall at intermediate
distances.

ChromaDepth 3-D is a viable stereoscopic display technology for
MirrorDome.  While relinquishing colour choice in generating content,
the programmer can produce a realistic depth effect that is observable
from all points within the dome and regardless of orientation of the
observers head.  Unlike conventional anaglyphs, the display looks sane
without glasses on, but when ChromaDepth 3-D glasses are worn, the
depth effect is quite remarkable.  

We have provided functionality in \splot\ to produce ChromaDepth 3-D
output on any device.  The function {\tt s2loadmap} can be used to
place the supplied ChromaDepth 3-D colourmap into \splot's colour
registers.  Then, function {\tt s2\-chro\-ma\-pts} or {\tt
s2\-chro\-ma\-cpts} can be used to plot points in 3D space, coloured
according to their distance from the camera.  In this way, we have
viewed the RAVE DR1 data in the MirrorDome.  By programming \splot\ to
colour the points --- using the correct ChromaDepth colour palette ---
according to velocity, coherent structure in the dataset was 
easily seen in the 3D immersive environment.  The combination of
MirrorDome and \splot's ChromaDepth 3-D capability will give an
extraordinary view of the survey once it nears completion.

\section{Future Possibilities}

The brief examples we have given of \splot\ programs only touch the
surface of what is possible.  The main motivation for developing
\splot\ was to provide a simple and effective entry-point to 3D
graphics for professional astronomers and astrophysicists.  However,
as a high-level programming library, \splot\ is genuinely flexible,
and we envisage a number of further domains in which it might be
useful.

{\em Public outreach.}  \splot\ might be used to
construct simple models of planetary systems, with the user
controlling visual effects such as which planets are displayed and
whether their orbital paths are drawn, or more physical parameters
such as planet and stellar masses and rotation periods.  The user
could easily stop and start the simulation, and move to pre-programmed
view points to demonstrate tilted orbits, for example.  If a
MirrorDome system is available, ``immersion'' in the simulated
planetary system is simple and effective, and concepts such as
retrograde motion can easily be explored.  Placing a panoramic sky
image ``behind'' the simulated system (as a texture on the interior of
a large-radius sphere) strengthens the immersive feeling and can make
\splot\ educational outreach programs exciting and thrilling for young
audiences.

{\em Professional presentation tools.}  With the ability to
place textures on planes and control animation states it is entirely
feasible to write code that uses the \splot\ library to support a
presentation at a conference.  Standard Microsoft PowerPoint content
can be saved to files for display within an \splot\ program, alongside
genuine realtime simulations of the concepts being discussed, or for
that matter dynamic display of realtime data streams.

{\em Real-time instrument monitoring.}  With its callback
mechanism, \splot\ can be operated in a mode where the main program
does nothing more than open the \splot\ device and register a callback
function.  This means that real-time instrument
monitoring programs can be written with \splot.  A simple example
might entail a callback function that monitors one or more disk files,
and creates geometry on each refresh cycle to reflect changes in the
files.  A more sophisticated callback function could use sockets to
connect to other processes, such as image acquisition software,
processing routines, or even environmental monitoring applications.

A concrete example is the potential to reimplement the Livedata
Monitor client that presents Parkes Multibeam data in a
waterfall-style display.  This client could be programmed to display
surface-elevation plots of time-frequency data using \splot.
Inspecting these plots from different viewing angles might offer new
possibilities for recognising patterns in radio-frequency interference
(RFI).  Stacking thirteen plots --- one on top of the other --- in 3D
space might also aid in recognising RFI that is impacting all feeds of
the system.  Or, for complex cross-correlation data, \splot\ makes
time-based plots of one complex variable possible without showing
phase and magnitude separately.  A product that is steadily rotating
in phase, with increasing amplitude, can be plotted as a widening
spiral in 3D space.

\section{Closing Remarks}
\label{sec:distribution}

We are making \splot\ available to the astronomical community.  It is
available as a binary library for GNU/Linux and Apple/OSX systems, and is
distributed with various sample programs (source code and pre-compiled
binaries) as a learning aid.
New users can make quick progress by adapting one of the sample
programs to their needs and can produce an executable program using the 
build scripts that are provided with the distribution.  Accelerated OpenGL 
performance should be
available by default on Apple/OSX systems, and on GNU/Linux systems
where accelerated NVidia drivers have been installed.  Even without
OpenGL-compliant hardware, GNU/Linux systems typically use the Mesa
library to provide software-mode rendering.  \splot\ requires an installation 
of the XForms\footnote{\url{http://world.std.com/~xforms/}} library.  Many 
distributions of GNU/Linux include XForms, and it 
is just a matter of selecting it for installation.  Apple/OSX users
can obtain a copy of XForms via 
Fink\footnote{\url{http://fink.sourceforge.net}}. 

\splot\ can be obtained from the following web page:
\begin{quote}
{\tt http://astronomy.swin.edu.au/s2plot}
\end{quote}
which contains information on installation, and detailed descriptions 
of all \splot\ functions.



\section*{Acknowledgments} 

We gratefully acknowledge Tim Pearson, for his PGPLOT Graphics
Subroutine Library, a much-loved package that deserves its place in
any collection of astronomical software.  PGPLOT's simple, clear
interface was an inspiration for many of the S2PLOT functions.  We
also offer our sincere thanks to Tony Fairall for alerting us to the
possibility of producing a chromostereoscopic display in a digital
dome environment.  We thank the following users for providing examples
of \splot\ use: Katie Kern, Alina Kiessling, Virginia Kilborn, Baerbel
Koribalski, Chris Power and Ashley Rowlands. We are also very grateful
to the referee for providing valuable suggestions on improvements to this 
paper.

\splot\ is freely available for non-commercial and educational use, however, 
it is not public-domain software.  The programming libraries and documentation
may not be redistributed or sub-licensed in any form without permission
from Swinburne University of Technology.  The programming library and
sample codes are provided without warranty.


\end{document}